# Bulk Superconductivity Induced by In-plane Chemical Pressure Effect in Eu$_{0.5}$La$_{0.5}$FBiS$_{2-x}$Se$_x$

Gen Jinno[1], Rajveer Jha[2], Akira Yamada[2], Ryuji Higashinaka[2], Tatsuma D. Matsuda[2], Yuji Aoki[2], Masanori Nagao[3], Osuke Miura[1], Yoshikazu Mizugchi[1*]

[1]*Department of Electrical and Electronic Engineering, Tokyo Metropolitan University, Hachioji 192-0397, Japan*

[2]*Department of Physics, Tokyo Metropolitan University, Hachioji 192-0397, Japan*

[3]*Center for Crystal Science and Technology, University of Yamanashi, Kofu 400-8511, Japan*

We have investigated Se substitution effect to superconductivity of an optimally-doped BiS$_2$-based superconductor Eu$_{0.5}$La$_{0.5}$FBiS$_2$. Eu$_{0.5}$La$_{0.5}$FBiS$_{2-x}$Se$_x$ samples with $x$ = 0-1 were synthesized. With increasing $x$, in-plane chemical pressure is enhanced. For $x \geq 0.6$, superconducting transitions with a large shielding volume fraction are observed in magnetic susceptibility measurements, and the highest $T_c$ is 3.8 K for $x$ = 0.8. From low-temperature electrical resistivity measurements, a zero-resistivity state is observed for all the samples, and the highest $T_c$ is observed for $x$ = 0.8. With increasing Se concentration, characteristics of electrical resistivity changes from semiconducting-like to metallic, suggesting that the emergence of bulk superconductivity is linked with the enhanced metallicity. A superconductivity phase diagram of the Eu$_{0.5}$La$_{0.5}$FBiS$_{2-x}$Se$_x$ superconductor is established. Temperature dependences of electrical resistivity show an anomalous two-step transition under high magnetic fields. Hence, the resistivity data are analyzed with assuming in-plane anisotropy of upper critical field.

1. Introduction

BiS$_2$-based superconductors have been drawing much attention as a new layered superconductor family [1-3]. The crystal structure is composed of alternate stacks of electrically conducting BiS$_2$ layers and insulating (blocking) layers, which resembles those of the Cu-based and the FeAs-based high transition temperature (high $T_c$) superconductors [4,5]. Without carrier doping, BiS$_2$-based compounds are a semiconductor with a band gap [1,6,7]. However, BiS$_2$-based compounds become metal (sometimes bad-metal) by introduction of electron carriers in the BiS$_2$ layers, and superconductivity is observed in electron-doped compounds. The scenario that superconductivity is induced in a band insulator by carrier doping seems conventional. However, the emergence of superconductivity in $BiS_2$-based compounds is related with not only carrier doping but also crystal structure optimization. The best example is $LaO_{1-x}F_xBiS_2$. Although superconducting transitions are observed in $LaO_{1-x}F_xBiS_2$ when electron carriers were doped by substitution of F for O, bulk superconductivity never appears while filamentary superconductivity with a small shielding volume fraction is observed. In addition, $T_c$ in the filamentary superconducting phase is less than 3 K for optimally doped $x = 0.5$ [2]. The absence of bulk superconductivity can be explained with the long in-plane Bi-S distance in $LaO_{1-x}F_xBiS_2$, which interferes the enhancement of orbital overlaps of in-plane Bi and S. This scenario can be linked to semiconducting-like behavior in the temperature dependence of electrical resistivity in $LaO_{1-x}F_xBiS_2$ with enough carrier doping. Therefore, to induce bulk superconductivity in $LaO_{1-x}F_xBiS_2$, the enhancement of orbital overlaps are required [8]; hereafter, we focus on $LaO_{0.5}F_{0.5}BiS_2$ ($x = 0.5$) to emphasize the importance of crystal structure to the emergence of bulk superconductivity.

There are two kinds of way to achieve the enhancement of orbital overlaps and bulk superconductivity. One of the strategies is *high pressure effect* [9]. $T_c$ of $LaO_{0.5}F_{0.5}BiS_2$ is largely enhanced from ~3 K to ~10 K under external high pressure above 1 GPa [9-12]. The large increase of $T_c$ under high pressure in $LaO_{0.5}F_{0.5}BiS_2$ accompanies with a structural transition from tetragonal (*P4/nmm*) to monoclinic (*P2$_1$/m*) [10]. In a monoclinic phase, a shorter Bi-S bond is showing up, with which the enhancement of orbital overlaps between Bi and S can be expected. In addition, structural distortion can be introduced by synthesizing samples under high pressure, as well as the external pressure effect, in $BiS_2$-based compounds [2,13-17]. The high-pressure synthesized $LaO_{0.5}F_{0.5}BiS_2$ samples contain local in-plane distortion [18] and show a high $T_c$, which is close to $T_c$ observed in high-pressure experiment. Namely, both application of external pressure and high-pressure synthesis can induce superconductivity in $LaO_{0.5}F_{0.5}BiS_2$.

The other strategy to induce bulk superconductivity in $LaO_{0.5}F_{0.5}BiS_2$ is *chemical pressure effect*. For example, isovalent substitution of $La^{3+}$ by smaller $Pr^{3+}$, $Nd^{3+}$, or $Sm^{3+}$ induces bulk superconductivity, and $T_c$ reaches above 5 K in $NdO_{0.5}F_{0.5}BiS_2$ or $Nd_{1-x}Sm_xO_{0.5}F_{0.5}BiS_2$ [19-22]. The rare earth (RE) substitution does not affect the structural symmetry (within X-ray diffraction) but compresses BiS planes. Compressing Bi-S distance enhances Bi-S orbital overlaps and then achieves bulk superconductivity. Similar chemical pressure effect can be produced in Se-substituted system, $LaO_{0.5}F_{0.5}BiS_{2-x}Se_x$ [23]. Although Bi-Ch (Ch: chalcogen) plane expands by substitution of in-plane $S^{2-}$ by larger $Se^{2-}$ [24], the packing density of Bi-Ch plane is enhanced due to the unchanged LaO blocking layer structure [8]. As a result, orbital overlaps of in-plane Bi and Ch is enhanced by Se substitution, and bulk superconductivity is induced by chemical pressure effect in $LaO_{0.5}F_{0.5}BiS_{2-x}Se_x$.

On the basis of these experimental facts, we recently proposed the importance of in-plane chemical pressure to the emergence of superconductivity in $REO_{0.5}F_{0.5}BiCh_2$ [8]. Both chemical pressure effects, such as RE-substitution and Ch-substitution effects in $REO_{0.5}F_{0.5}BiCh_2$, can be regarded as the same effect from the viewpoint of the tuning of orbital overlaps of Bi and Ch. In addition, this scenario could be extensible to the pressure effect. Therefore, it is important to obtain further evidences on this scenario, and hence, investigation of in-plane chemical pressure effect to superconductivity (particularly Se-substitution effect) in new system, which is different from $REO_{0.5}F_{0.5}BiCh_2$ but has a similar situation (similar lattice structure and physical properties) with $LaO_{0.5}F_{0.5}BiS_2$. A good candidate is RE-substituted $EuFBiS_2$, $Eu_{0.5}RE_{0.5}FBiS_2$ [25,26]. The substitution of $Eu^{2+}$ by $RE^{3+}$ introduces 0.5 electrons per Bi, which is the same amount as $LaO_{0.5}F_{0.5}BiS_2$. The lattice constant of $Eu_{0.5}RE_{0.5}FBiS_2$ is $a \sim 4.07$ Å, which is closed to that of $LaO_{0.5}F_{0.5}BiS_2$. In addition, $Eu_{0.5}La_{0.5}FBiS_2$ shows filamentary superconductivity with an onset $T_c$ ($T_c^{onset}$) of ~2.3 K, and $T_c$ increases up to 10 K under high pressure [26]. Considering these quite similar situations in between $Eu_{0.5}La_{0.5}FBiS_2$ and $LaO_{0.5}F_{0.5}BiS_2$, we expected the emergence of bulk superconductivity in Se-substituted $Eu_{0.5}La_{0.5}FBiS_{2-x}Se_x$. Here, we report the emergence of superconductivity by Se substitution in $Eu_{0.5}La_{0.5}FBiS_{2-x}Se_x$, which provides new evidence that in-plane chemical pressure effect can induce bulk superconductivity in $BiCh_2$-based compounds.

2. Experimental details

Polycrystalline samples of $Eu_{0.5}La_{0.5}FBiS_{2-x}Se_x$ with $x$ = 0, 0.2, 0.4, 0.6, 0.8, and 1 were prepared by a solid state reaction method. Powders of EuS (99.9%), $La_2S_3$ (99.9%), $BiF_3$ (99.9%), and $LaF_3$ (99.9%) and grains of Bi (99.999%), S (99.99%) and Se (99.999%) were mixed, pelletized, and sealed in an evacuated quartz tube. The heating temperature was optimized for each composition: the samples were heated for 20 h at 780ºC for $x$ = 0 and 0.2, at 700ºC for $x$ = 0.4 and 0.6, and at 650ºC for $x$ = 0.8 and 1. The obtained sample was ground, mixed, pelletized, and heated with the same heating condition.

The obtained samples were characterized using powder X-ray diffraction (XRD) with a CuKa radiation by the $\theta$-$2\theta$ method. The XRD patterns were analyzed using the Rietveld method [27]. Temperature ($T$) dependence of electrical resistivity ($\rho$) under magnetic fields was measured using a four-terminal method with Physical Property Measurement System (PPMS, Quantum Design) down to $T$ = 0.5 K. Temperature dependence of magnetic susceptibility ($\chi$) was measured using a superconducting quantum interference devise (SQUID) magnetometer with Magnetic Property Measurement System (MPMS-3) by the SQUID-VSM mode.

3. Results and discussion

Figure 1(a) shows the XRD patterns for $x$ = 0-1. The XRD patterns were analyzed by a two-phase analysis for $x$ = 0-0.6 and three-phase analysis for $x$ = 0.8 and 1. The fitting results are shown in Figs. S1-S6 (Supplementary Materials [28]). Obtained structural parameters and refined populations of the impurity phases are listed in Tables S1 and S2 (Supplementary Materials [28]). The major phase is characterized as the EuFBiS$_2$-type tetragonal phase with space group of $P4/nmm$ [29]. For $x$ = 0-0.6, small amount of BiF$_3$ impurity (less than 5%) was detected. For $x$ = 0.8 and 1, Bi$_2$Se$_3$ impurity (~12%) was detected. Figures 1(b) and 1(c) show the $x$ dependence of lattice constant ($a$ and $c$). The lattice constant of $a$ monotonically increases with increasing $x$, which can be understandable with the difference of ionic radius of Se$^{2-}$ (198 pm, assuming a coordination number of 6) and S$^{2-}$ (184 pm). In contrast, the lattice constant of $c$ does not show an apparent expansion with increasing $x$. These evolutions of the lattice constants by Se substitution are similar to those observed in LaO$_{0.5}$F$_{0.5}$BiS$_{2-x}$Se$_x$ [23]. The anisotropic expansion along the $a$-axis in LaO$_{0.5}$F$_{0.5}$BiS$_{2-x}$Se$_x$ is resulting from the selective substitution of the in-plane S1 site by Se [24]. In this study, we tried to refine the Se occupancy for the Ch1 and Ch2 sites, but Se occupancy at Ch2 was refined as almost zero or a small negative value for $x$ = 0-0.6. Hence, we fixed Se occupancy at Ch1 as nominal $x$ = 0-0.6. For $x$ = 0.8 and 1, small amount of Se was refined at Ch2 site. These results suggest the doped Se selectively occupies the in-plane Ch1 site. Although further analysis with synchrotron XRD or single crystal XRD is needed to precisely determine the Se occupancy, the tendency that doped Se occupies the in-plane Ch1 site is the same as LaO$_{0.5}$F$_{0.5}$BiS$_{2-x}$Se$_x$.

As demonstrated in Ref. 8, in-plane chemical pressure (CP), which is calculated as CP = ($R_{Bi}$ + $R_{Ch}$) / (Bi-Ch bond distance), where $R_{Bi}$ and $R_{Ch}$ are ionic radii of Bi and Ch, can be a good scale to qualitatively compare the magnitude of in-plane CP and physical properties in BiCh$_2$ compounds [8]. We used $R_{Bi}$ estimated from average Bi-S bond distances of $LaO_{0.54}F_{0.46}BiS_2$, as demonstrated in Ref. 8 [30]. Figure 2 shows the $x$ dependences of in-plane CP for $Eu_{0.5}La_{0.5}FBiS_{2-x}Se_x$ and $LaO_{0.5}F_{0.5}BiS_{2-x}Se_x$. We estimated Bi-Ch bond distance using the structural parameters obtained from Rietveld refinement. It is clear that the evolution of in-plane CP in $Eu_{0.5}La_{0.5}FBiS_{2-x}Se_x$ is almost the same as that in $LaO_{0.5}F_{0.5}BiS_{2-x}Se_x$. Therefore, we can expect the emergence of bulk superconductivity in $Eu_{0.5}La_{0.5}FBiS_{2-x}Se_x$, as well as in $LaO_{0.5}F_{0.5}BiS_{2-x}Se_x$.

Figure 3 shows the $T$ dependences of $\chi$ for $x = 0$-1, which were measured after zero-field cooling (ZFC). Although clear superconducting signals are not observed for $x = 0$ and 0.2, diamagnetic signals corresponding to the emergence of superconductivity are observed for $x = 0.4$ -1. The shielding volume fraction increases with increasing $x$, and $\Delta\chi$ exceeds $-1/4\pi$ for $x = 0.6, 0.8$, and 1. Indeed, Se substitution can induce bulk superconductivity in $Eu_{0.5}La_{0.5}FBiS_{2-x}Se_x$ at $x \geq 0.6$. $T_c$ increases with increasing $x$ and achieves $T_c = 3.7$ K for $x = 0.8$, and $T_c$ of $x = 1$ is slightly lower than $T_c$ of $x = 0.8$. In addition, we measured $\chi$-$T$ up to 300K to investigate the normal state magnetism. There is no magnetic transition except for the superconducting transition, as shown in Fig. S7 (Supplementary Materials [28]), which shows $T$ dependence of $\chi$ for $x = 0.4$ at $T < 300$ K.

Figure 4(a) shows the $T$ dependences of $\rho$ for $x = 0$-1. For $x = 0$ and 0.2, $\rho$ largely increases with decreasing $T$, which is semiconducting-like behavior and implying bad electron hopping in conduction layers. With increasing $x$, $\rho$ is strongly suppressed, and metallic-like conductivity is induced for $x = 0.8$ and 1, as displayed in Fig. S9(a) (Supplementary Materials [28]). The enhanced metallicity in Se-doped compounds is consistent with the enhanced in-plane CP (Fig. 2), which is quite similar to the case of $LaO_{0.5}F_{0.5}BiS_{2-x}Se_x$ [23].

On superconductivity characteristics, $T_c$ shows nonlinear evolution at a low-$x$ regime. The inset figure of Fig. 4(a) shows $\rho$-$T$ curves for $x = 0$-0.4 at low temperatures. For $x = 0$, $T_c^{onset}$ is 2.2 K, and zero resistivity is observed below $T_c^{zero} = 1.2$ K. Except for $x = 0.2$ (discussed later), $T_c$ increases with increasing $x$ in $Eu_{0.5}La_{0.5}FBiS_{2-x}Se_x$. The superconducting transitions for $x = 0.6, 0.8$, and 1 are enlarged in Fig. 4(b). The highest $T_c$ is observed in $x = 0.8$. $T_c^{zero}$ is 3.8 K, which is almost the same as $T_c$ estimated from $\chi$-$T$ for $x = 0.8$. Interestingly, $\rho$ for those superconducting samples begins to decrease at a temperature well above $T_c^{zero}$: $T_c^{onset}$ is estimated to be 6.2 K for $x = 0.8$ (Fig. S9(b) (Supplementary Materials [28])). Typically, high $T_c^{onset}$ well above $T_c^{zero}$ (bulk $T_c$) has been observed in several $BiS_2$-based superconductors. For example, $NdO_{0.5}F_{0.5}BiS_2$ single crystals show a $T_c^{onset}$ of ~20 K and a bulk $T_c$ of 4-5 K [31]. Therefore, the drop of $\rho$ at 6.2 K in $x = 0.8$ can be regarded as a local onset of superconductivity.

Then, we consider the suppression of superconductivity by 20%-Se substitution (in $x = 0.2$). Although $\rho$-$T$ of $x = 0.2$ exhibits $T_c^{onset} = 2.7$ K, $\rho$ does not drop to zero and shows a second superconducting transition with $T_c^{onset} = 1.3$ K and $T_c^{zero} = 0.5$ K in $x = 0.2$. The two-step transition should be caused by the existence of minor regions having a higher $T_c$ of 2.7 K, and $T_c$ of the major phase should be lower than that of $x = 0$, because of $T_c^{zero} = 0.5$ K. The decrease of bulk $T_c$ by 20%-Se substitution is also seen in $LaO_{0.5}F_{0.5}BiS_{2-x}Se_x$ [23]. The suppression of $T_c$ may be explained by a local distortion scenario. We assume that superconductivity observed in $x = 0$ is not bulk in nature and emerging due to local distortion (or disorder). In other words, the tetragonal $Eu_{0.5}La_{0.5}FBiS_2$ ($x = 0$) should not be a bulk superconductor due to low in-plane CP. In general, the crystal structure of $BiCh_2$-based superconductors is flexible in between tetragonal (P4/$nmm$) and monoclinic ($P2_1/m$) [32], and the in-plane structure contains large disorder [18,33-35]. In fact, $LaOBiS_2$-based and $EuFBiS_2$-based superconductors show a large increase of $T_c$ under high pressure, which accompanies a structural transition from tetragonal to monoclinic [9-12,36-37]. Assuming the presence of local distortion, which induces filamentary superconductivity in $x = 0$, we consider that the superconductivity with $T_c^{zero} = 0.5$ is induced in the tetragonal phase of $x = 0.2$ when the local in-plane distortion (or disorder) is suppressed by the enhancement of in-plane CP, as observed in $Ce_{1-x}Nd_xO_{0.5}F_{0.5}BiS_2$ [35].

From the $\chi$-$T$ and the $\rho$-$T$ measurements, a superconductivity phase diagram is established (Fig. 5). For $x = 0$ and 0.2, a semiconducting-like behavior is observed in $\rho$-$T$, and bulk superconductivity is not observed. For $x = 0.4$, the semiconducting-like behavior in $\rho$-$T$ is suppressed, and superconducting transition is clearly observed in both $\rho$-$T$ and $\chi$-$T$. With increasing $x$, metallic conductivity is induced, and bulk superconductivity is induced; the highest $T_c$ is $T_c^{zero} = 3.8$ K for $x = 0.8$. Similarly to $LaO_{0.5}F_{0.5}BiS_{2-x}Se_x$, both metallic conductivity and bulk superconductivity are induced by Se substitution in $Eu_{0.5}La_{0.5}FBiS_{2-x}Se_x$ as well. We, thus, propose that the in-plane CP is one of the most important parameters for the emergence of metallic conductivity and bulk superconductivity in $BiCh_2$-based superconductors.

We discuss the electrical transport properties of $x = 0.6$ and 0.8 under high magnetic fields. The $x = 0.6$ sample shows bulk superconductivity and is qualified as almost single-phase in XRD, so that we can reliably obtain intrinsic superconducting properties from the transport properties under magnetic fields. The $x = 0.8$ sample contains 13% impurity phases, but the $T_c$ is the highest among the series. Figures 6(a) and 6(b) show $\rho$-$T$ under magnetic fields up to $\mu_0H = 5$ T for $x = 0.6$ and 0.8, respectively. $T_c^{zero}$ is suppressed with increasing $H$. The $T_c^{onset}$, for example, $T_c^{onset} = 6.2$ K for $x = 0.8$, is also suppressed with increasing $H$. Noticeably, at high $H$, other kink appears below $T_c^{onset}$, as if there are two kinds of $T_c^{onset}$. Similar behavior is sometimes observed in BiCh$_2$-based superconductors [2,26,31,38,39]. Namely, there are, at least, two components related to the emergence of superconductivity in Eu$_{0.5}$La$_{0.5}$FBiS$_{2-x}$Se$_x$; one component is relatively strong against the high magnetic field and the other is largely suppressed by the high magnetic field. In a previous study on upper critical field ($H_{c2}$) of LaO$_{0.5}$F$_{0.5}$BiS$_2$, $\rho$-$T$ under high magnetic fields was analyzed from temperature dependences of the $T$ derivative of $\rho$, d$\rho$/d$T$ [38]. With the $H_{c2}$ analysis based on $T$-derivative of $\rho$ or $\chi$, anisotropy of $H_{c2}$ has been estimated using a polycrystalline sample of several superconductors [38,40].

$T$ dependences of d$\rho$/d$T$ of Eu$_{0.5}$La$_{0.5}$FBiS$_{2-x}$Se$_x$ show two kinds of inflection points, as shown in Fig. 6(a) and Fig. S10 (Supplementary Materials [28]). Here, we define two kinds of $T_c^{onset}$ as $T_c^1$ and $T_c^2$; the estimation of $T_c^1$ and $T_c^2$ is described in Fig. S10. $T_c^1$ and $T_c^2$ estimated from d$\rho$/d$T$ correspond to the higher $T_c^{onset}$ and the lower $T_c^{onset}$ in $\rho$-$T$, respectively. For $x = 0.6$ and 0.8, all $T_c^1$ and $T_c^2$ are estimated in the same manner and plotted in Figs. 6(c) and 6(d); the data points are represented as upper critical fields, $H_{c2}^1$ and $H_{c2}^2$, in Figs. 6(c) and 6(d), respectively. In addition, irreversible field, $H_{irr}$, is estimated from $T_c^{zero}$ and plotted in Figs. 6(c) and (d). With decreasing $T$, $H_{c2}^1$ and $H_{c2}^2$ increase, but the robustness of superconductivity against $H$ is clearly different in between $H_{c2}^1$ and $H_{c2}^2$ for both $x$. The anomalous anisotropy of $H_{c2}$ cannot be explained by the anisotropy of $H_{c2}$ in between $H//a$ and $H//c$ because BiCh$_2$-based superconductors exhibit extremely high anisotropy of superconductivity with an anisotropy parameter ($\gamma$) of 30-40 [41]. Therefore, the presence of two kinds of $H_{c2}$, anisotropy of $H_{c2}$, should be regarded as in-plane anisotropy, as proposed for LaO$_{0.5}$F$_{0.5}$BiS$_2$ [38]. Although such in-plane anisotropy of superconductivity should not appear in a perfect tetragonal superconductor, in-plane local distortions can lift degeneracy of the Bi-6$p_x$ and Bi-6$p_y$ orbitals, which locally induces anisotropy of superconducting states in the BiCh plane. We mentioned possible suppression of in-plane distortion by Se substitution in above discussion of the $x$ dependence of $T_c^{zero}$, but it should be possible that the in-plane local distortion still exists and causes a local high-$T_c$ phase ($T_c^1$ and $H_{c2}^1$). We notice that the data points of $H_{irr}$ and $H_{c2}^2$ of $x = 0.8$ almost correspond to those of $x = 0.6$. This suggests that bulk superconductivity is almost induced at $x = 0.6$. Further Se substitutions, $x > 0.6$, do not largely affect the bulk properties of $Eu_{0.5}La_{0.5}FBiS_{2-x}Se_x$ but enhance $T_c^{onset}$ ($T_c^1$) and metallicity. To understand the origin of the higher $T_c^{onset}$ ($T_c^1$), further investigations are needed.

4. Conclusion

We synthesized a new $BiCh_2$-based superconductor system, $Eu_{0.5}La_{0.5}FBiS_{2-x}Se_x$, with the expecting of the emergence of bulk superconductivity by in-plane chemical pressure effect. Polycrystalline samples of $Eu_{0.5}La_{0.5}FBiS_{2-x}Se_x$ with $x = 0-1$ were synthesized by solid state reaction method. The XRD and Rietveld analyses revealed that Se selectively occupied at the in-plane Ch1 site, and hence, in-plane chemical pressure was enhanced. For $x \geq 0.6$, superconducting transitions with a large shielding volume fraction were observed in $\chi$-$T$ measurements, and the highest $T_c$ was 3.8 K for $x = 0.8$. From $\rho$-$T$ measurements, a zero-resistivity state was observed for all the samples, and the highest $T_c$ was observed for $x = 0.8$. With increasing Se concentration, characteristics of $\rho$-$T$ changed from semiconducting-like to metallic, suggesting that the emergence of bulk superconductivity is linked with the enhanced metallicity via in-plane chemical pressure. A superconductivity phase diagram of $Eu_{0.5}La_{0.5}FBiS_{2-x}Se_x$ superconductor was established. The $\rho$-$T$ data under high magnetic fields were analyzed with assuming in-plane anisotropy of upper critical field, which indicated local in-plane distortion may result in the high-$T_c^{onset}$, obviously higher than the bulk $T_c$.

**Acknowledgment**

This work was partly supported by Grants-in-Aid for Scientific Research (Nos. 15H05886, 15H05884, 25707031, 15H03693 and 16H04493).

*E-mail: mizugu@tmu.ac.jp

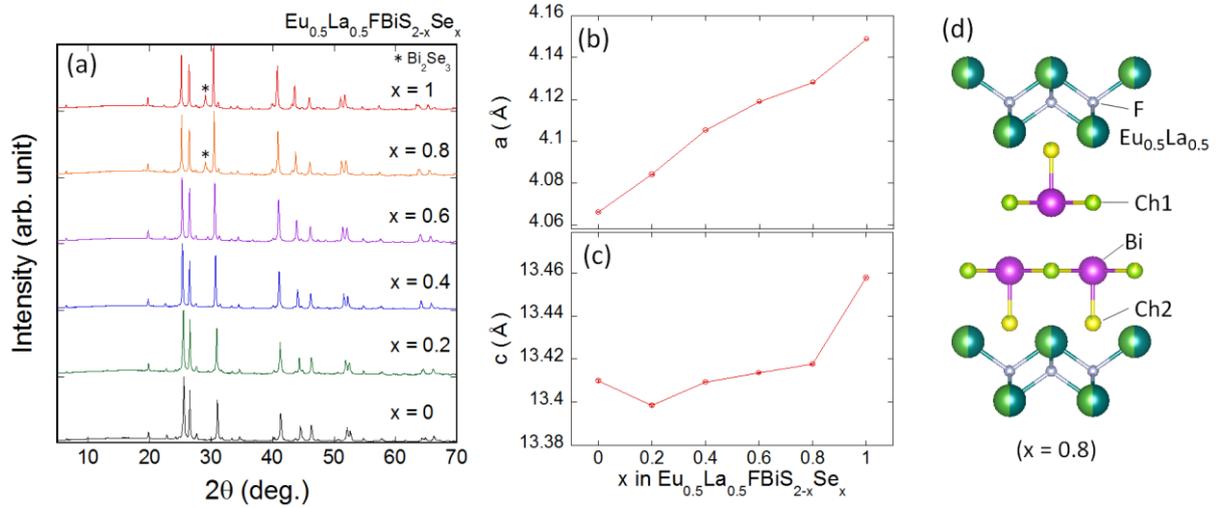

Fig. 1. (a) XRD patterns for Eu$_{0.5}$La$_{0.5}$FBiS$_{2-x}$Se$_x$ with $x$ = 0–1. Asterisks indicate peaks corresponding to a Bi$_2$Se$_3$ impurity. Rietveld fitting for each XRD pattern is shown in Supplementary Materials [28]. (b,c) Se concentration dependence of lattice constant ($a$ and $c$). (d) Schematic image of crystal structure of Eu$_{0.5}$La$_{0.5}$FBiS$_{2-x}$Se$_x$: here, $x$ = 0.8 is depicted as an example. Ch1 and Ch2 denote the in-plane chalcogen site and the out-of-plane chalcogen site. The crystal structure image was depicted using VESTA program [42].

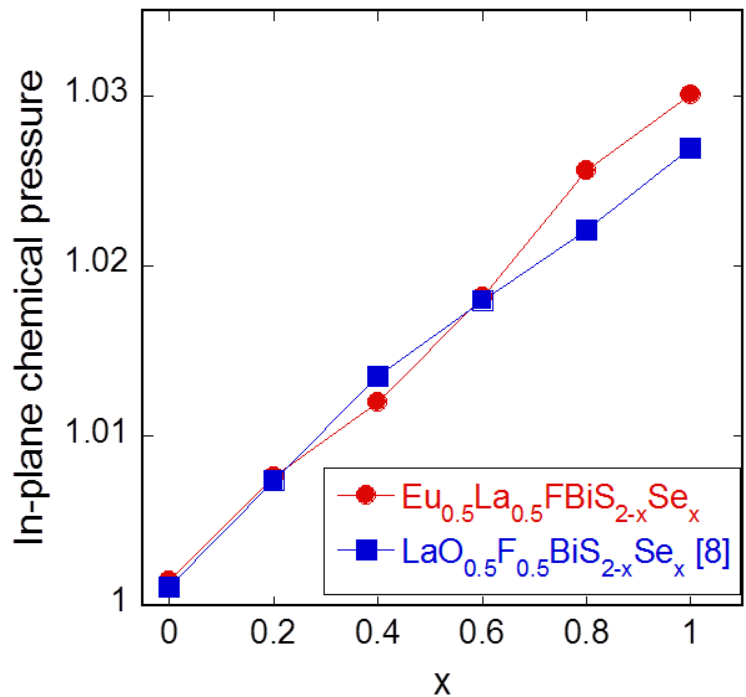

Fig. 2. Comparison of the evolution of in-plane chemical pressure by Se substitution in Eu$_{0.5}$La$_{0.5}$FBiS$_{2-x}$Se$_x$ and LaO$_{0.5}$F$_{0.5}$BiS$_{2-x}$Se$_x$.

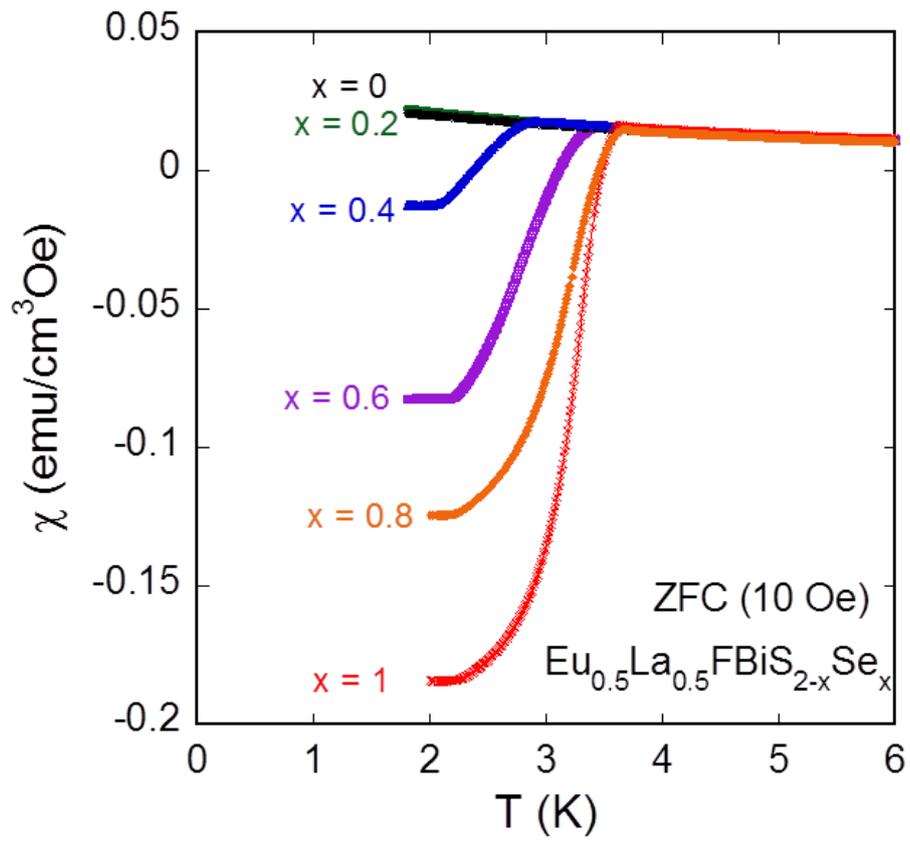

Fig. 3. Temperature ($T$) dependences of magnetic susceptibility ($\chi$) for $Eu_{0.5}La_{0.5}FBiS_{2-x}Se_x$ measured after zero-field cooling (ZFC).

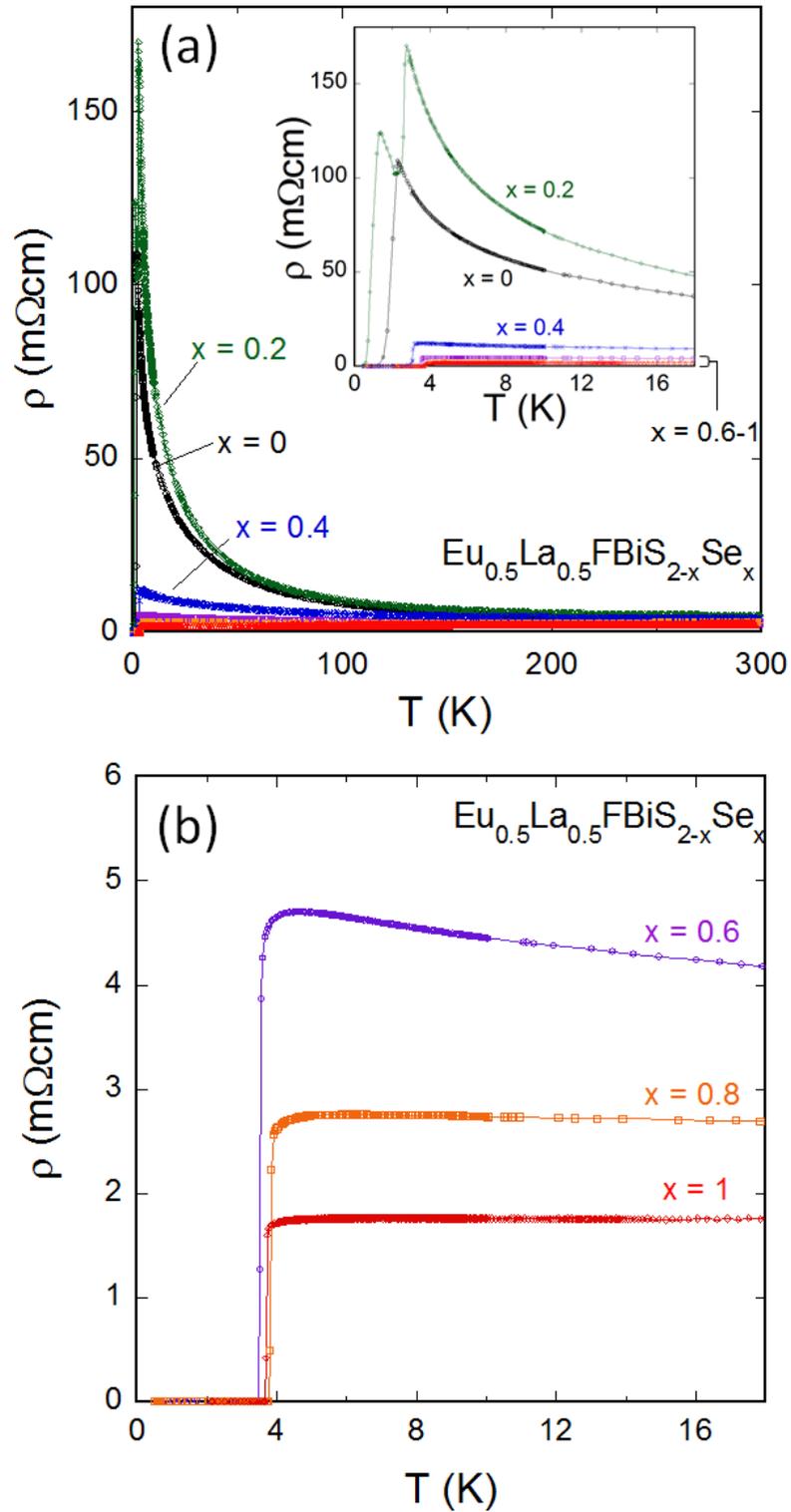

Fig. 4. (a) Temperature ($T$) dependences of electrical resistivity ($\rho$) for $Eu_{0.5}La_{0.5}FBiS_{2-x}Se_x$. In the inset, superconducting transitions for $x$ = 0–0.4 are enlarged. (b) $T$ dependences of $\rho$ for $x$ = 0.6–1 at low temperatures.

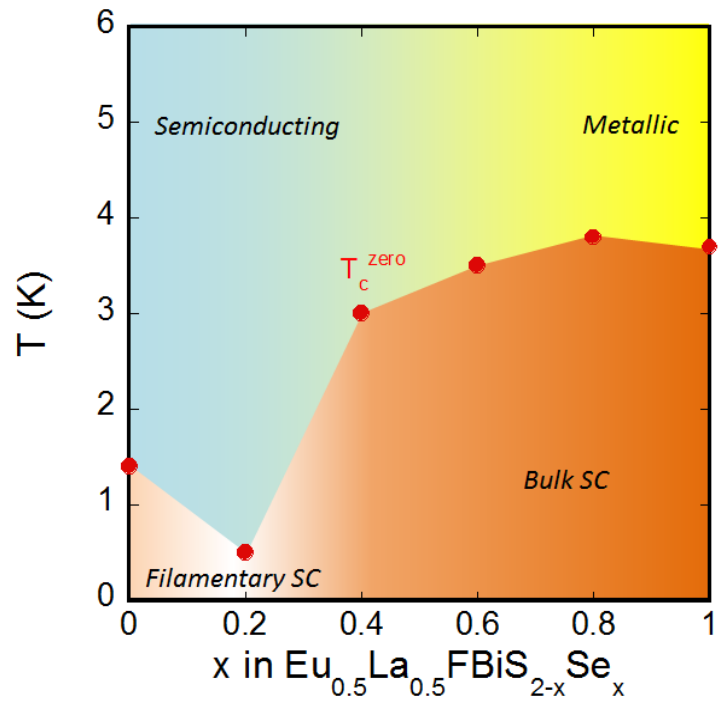

Fig. 5. Superconductivity phase diagram of $Eu_{0.5}La_{0.5}FBiS_{2-x}Se_x$ with $T_c^{zero}$ estimated from the $\rho$-$T$ data. SC denotes superconductivity.

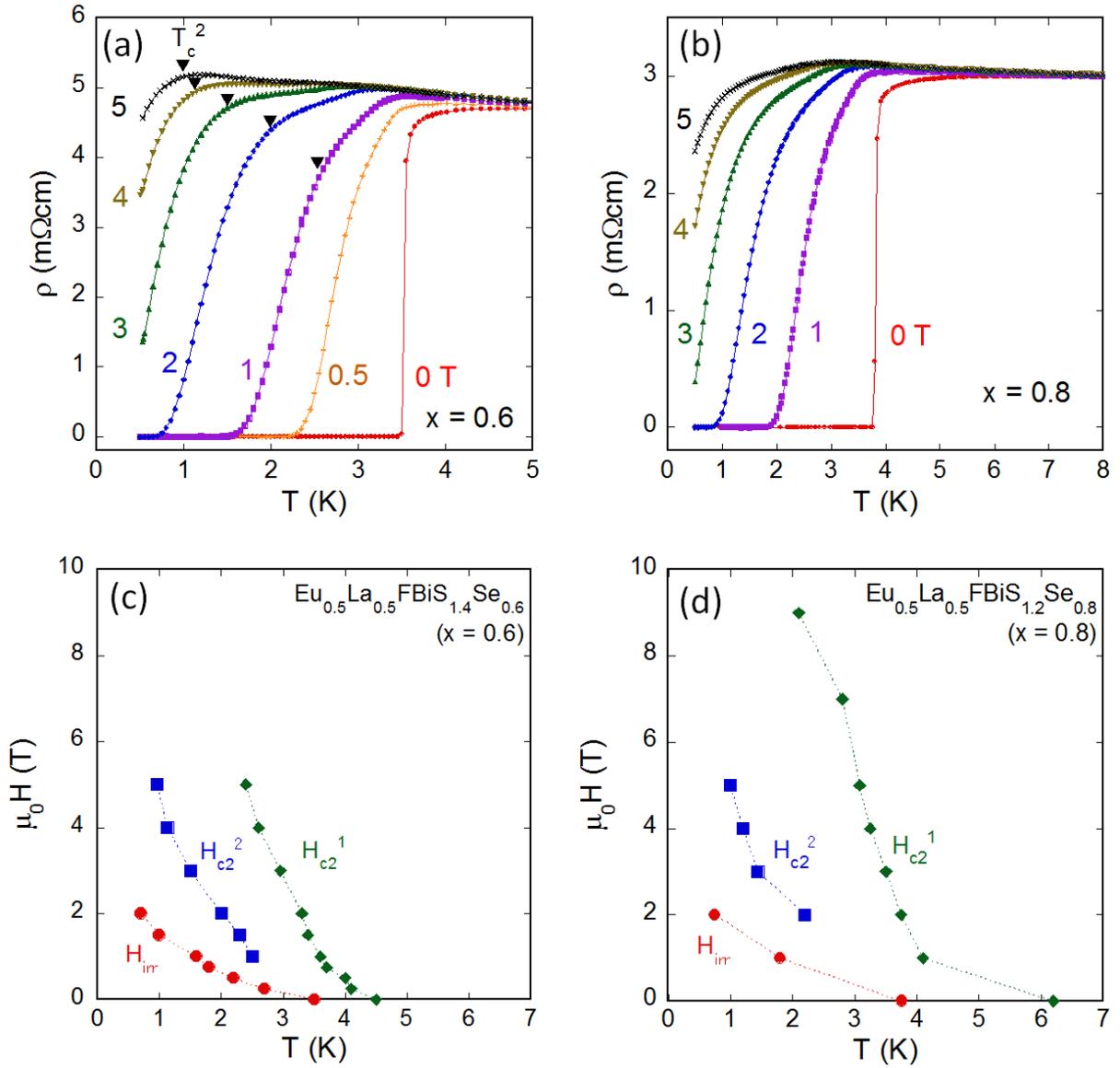

Fig. 6. (a,b) Temperature ($T$) dependences of electrical resistivity ($\rho$) for $x = 0.6$ (Eu$_{0.5}$La$_{0.5}$FBiS$_{1.4}$Se$_{0.6}$) and $x = 0.8$ (Eu$_{0.5}$La$_{0.5}$FBiS$_{1.2}$Se$_{0.8}$) under magnetic fields up to 5 T. The lower onset temperature, $T_c^2$, is indicated with reverse triangle in Fig. 6(a). (c,d) $H$-$T$ phase diagram of $x = 0.6$ and 0.8. Irreversible field, $H_{irr}$, is plotted with $T_c^{zero}$. Two kinds of upper critical fields, $H_{c2}^2$ and $H_{c2}^1$, are plotted with characteristic onset temperatures, $T_c^2$ and $T_c^1$.

# Supplemental Materials

Table S1. Refined structural parameters for $Eu_{0.5}La_{0.5}FBiS_{2-x}Se_x$. Ch1 and Ch2 sites are the in-plane and out-of-plane chalcogen sites. The atomic coordinates of $Eu_{0.5}La_{0.5}FBiS_{2-x}Se_x$ in the tetragonal P4/*nmm* space group can be described as (0, 0.5, $z$) for Eu/La, Bi, Ch1, and Ch2, and (0, 0, 0) for F. Thermal factor ($B$) for Eu/La, F, Bi, Ch1, Ch2 were fixed as 0.5, 1, 1.5, 1.5, and 1, which were typical values determined from synchrotron XRD in 1112-type compounds. In this study, we tried to refine the Se occupancy for the Ch1 and Ch2 sites, but Se occupancy at Ch2 was refined as almost zero or a small negative value for $x$ = 0-0.6. Hence, we fixed Se occupancy at Ch1 as nominal $x$ = 0-0.6. For $x$ = 0.8 and 1, small amount of Se was refined at Ch2 site: 1%-Se (for $x$ = 0.8) and 7%-Se (for $x$ = 1) were refined for Ch2 site.

| $x$ | $a$ (Å) | $c$ (Å) | $z$(Eu/La) | $z$(Bi) | $z$(Ch1) | $z$(Ch2) | $R_{wp}$ (%) |
|---|---|---|---|---|---|---|---|
| 0 | 4.0661(2) | 13.4080(5) | 0.1090(3) | 0.6247(2) | 0.366(2) | 0.827(1) | 8.5 |
| 0.2 | 4.0841(2) | 13.3984(5) | 0.1090(3) | 0.6252(2) | 0.375(1) | 0.8250(8) | 7.8 |
| 0.4 | 4.10520(7) | 13.4094(3) | 0.1091(2) | 0.6256(1) | 0.3764(5) | 0.8204(5) | 5.6 |
| 0.6 | 4.1188(1) | 13.4136(4) | 0.1091(2) | 0.6263(2) | 0.3779(6) | 0.8218(7) | 7.6 |
| 0.8 | 4.1280(1) | 13.4177(4) | 0.1098(3) | 0.6259(2) | 0.3760(8) | 0.8216(8) | 6.5 |
| 1 | 4.1488(1) | 13.4577(3) | 0.1095(3) | 0.6268(2) | 0.3735(7) | 0.8218(7) | 6.1 |

Table S2. Refined populations of the $Eu_{0.5}La_{0.5}FBiS_{2-x}Se_x$ phase and impurity phases.

| $x$ | $Eu_{0.5}La_{0.5}FBiS_{2-x}Se_x$ | $BiF_3$ | $Bi_2Se_3$ |
|---|---|---|---|
| 0 | 95% | 5% | - |
| 0.2 | 96% | 4% | - |
| 0.4 | 97% | 3% | - |
| 0.6 | 97% | 3% | - |
| 0.8 | 87% | 1% | 12% |
| 1 | 87% | 1% | 12% |

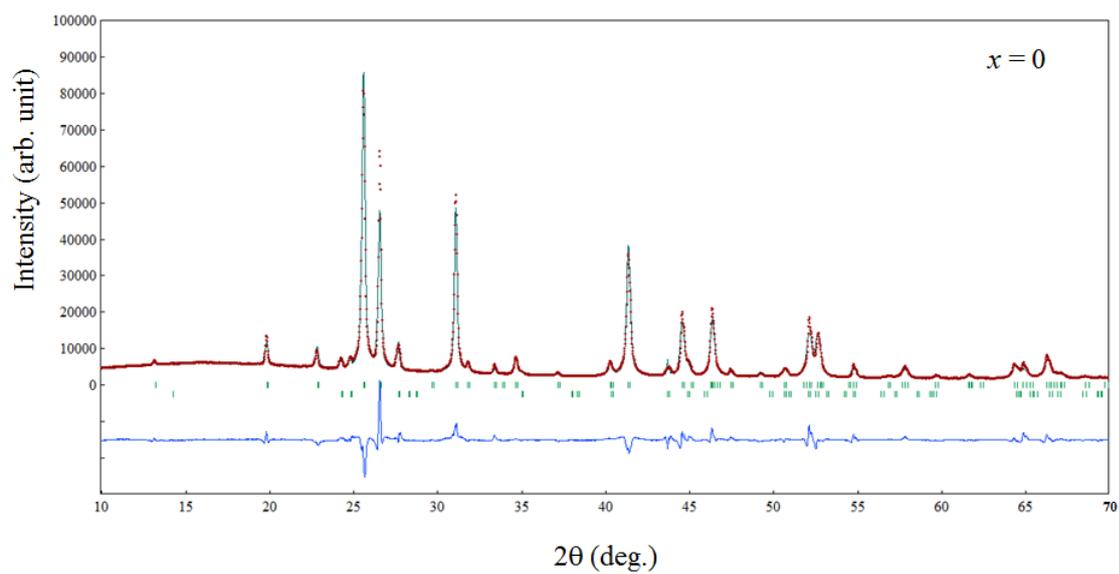

Fig. S1. XRD pattern with Rietveld fitting for $x = 0$. The second phase used in Rietveld refinement is $BiF_3$.

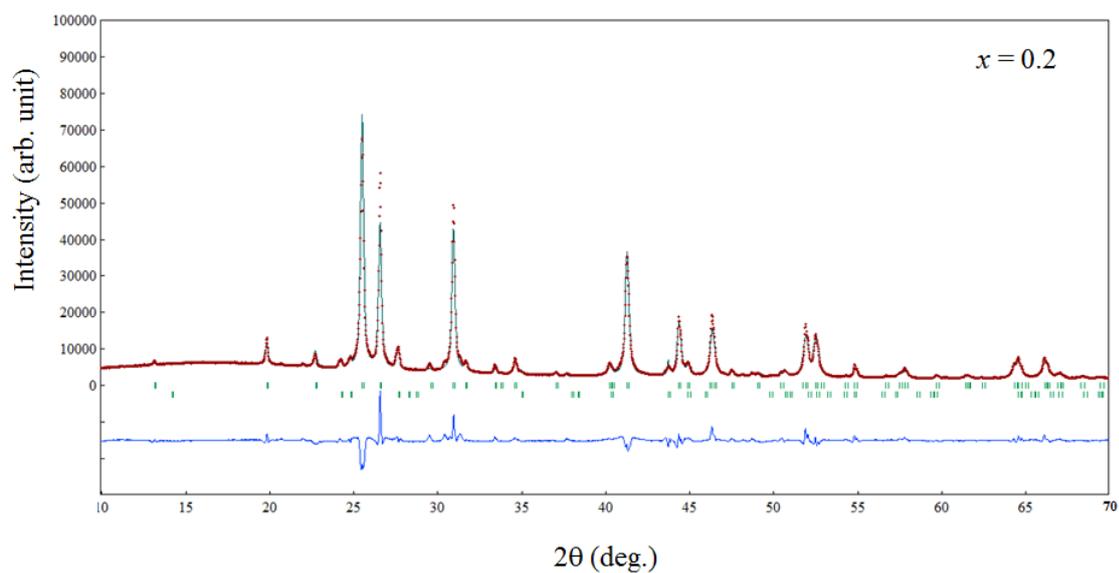

Fig. S2. XRD pattern with Rietveld fitting for $x = 0.2$. The second phase used in Rietveld refinement is $BiF_3$.

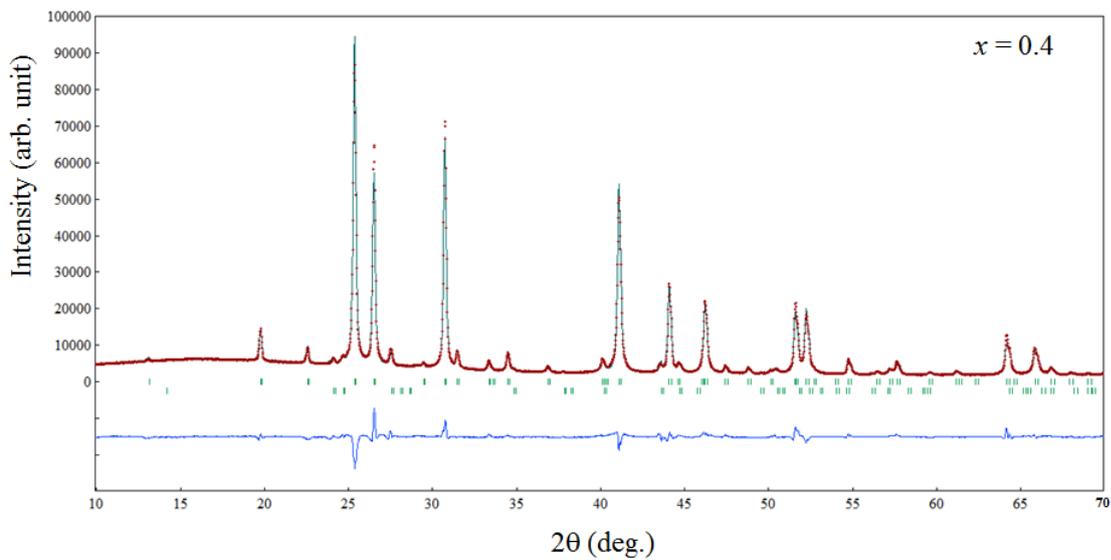

Fig. S3. XRD pattern with Rietveld fitting for $x = 0.4$. The second phase used in Rietveld refinement is $BiF_3$.

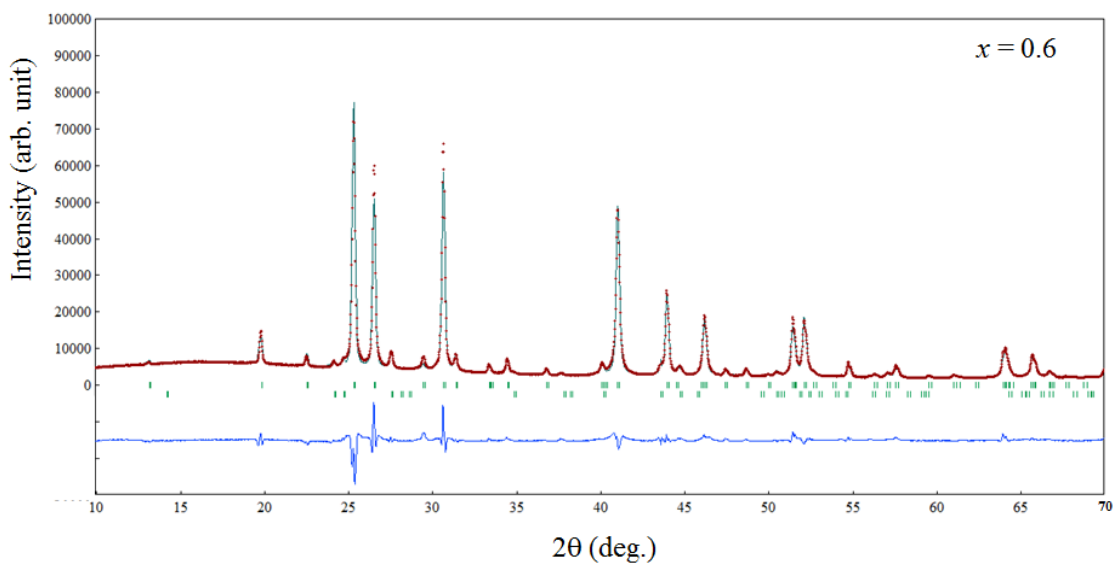

Fig. S4. XRD pattern with Rietveld fitting for $x = 0.6$. The second phase used in Rietveld refinement is $BiF_3$.

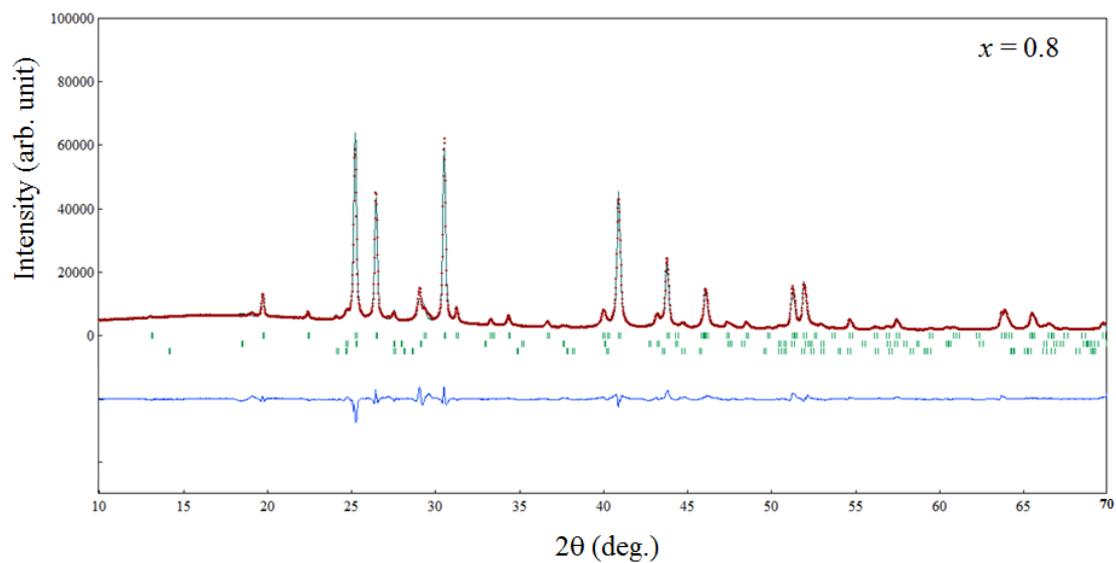

Fig. S5. XRD pattern with Rietveld fitting for $x = 0.8$. The second and third phases used in Rietveld refinement are $Bi_2Se_3$ and $BiF_3$.

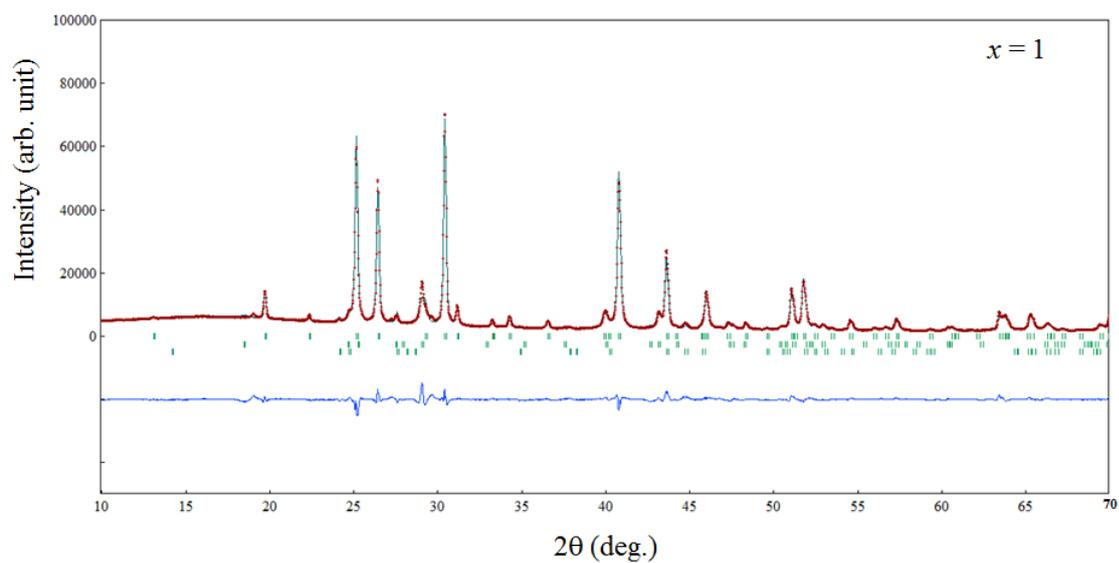

Fig. S6. XRD pattern with Rietveld fitting for $x = 1$. The second and third phases used in Rietveld refinement are $Bi_2Se_3$ and $BiF_3$.

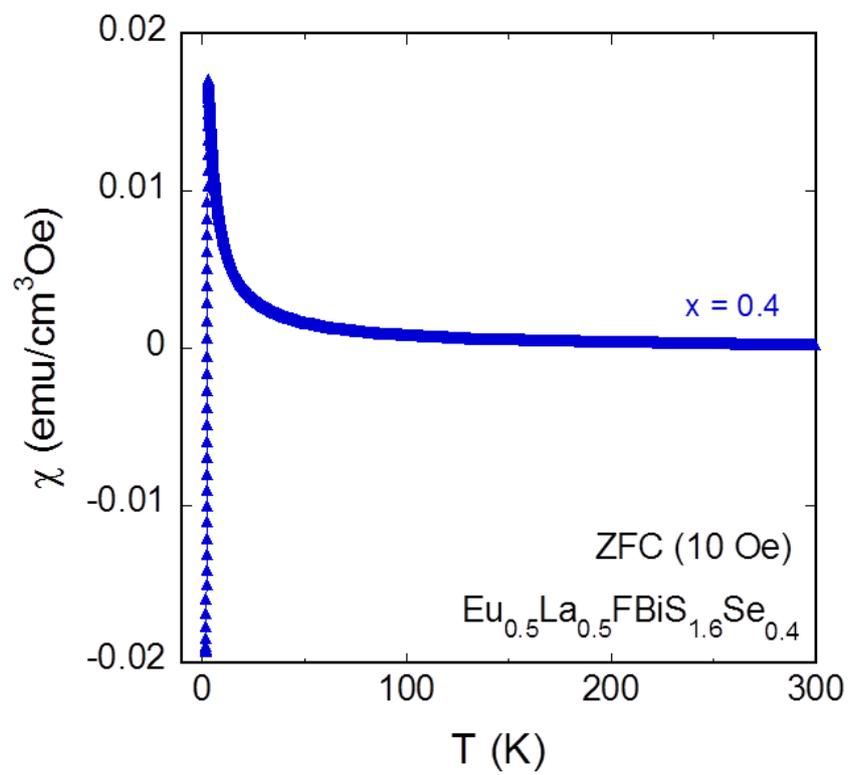

Fig. S7. Temperature dependence of magnetic susceptibility ($\chi$) for $x$ = 0.4 ($Eu_{0.5}La_{0.5}FBiS_{1.6}Se_{0.4}$).

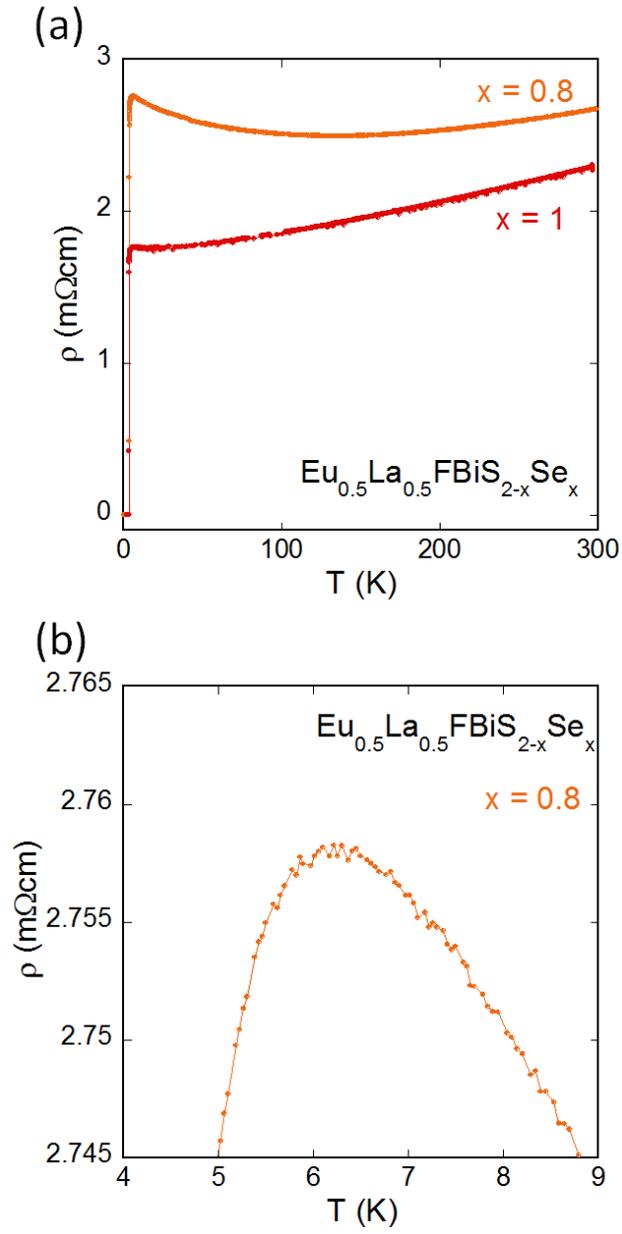

Fig. S9. (a) Temperature dependences of electrical resistivity ($\rho$) for $x = 0.8$ and 1. (b) Enlargement of $\rho$-$T$ for $x = 0.8$ at around $T_c^{onset}$, which will be defined as $T_c^1$ later.

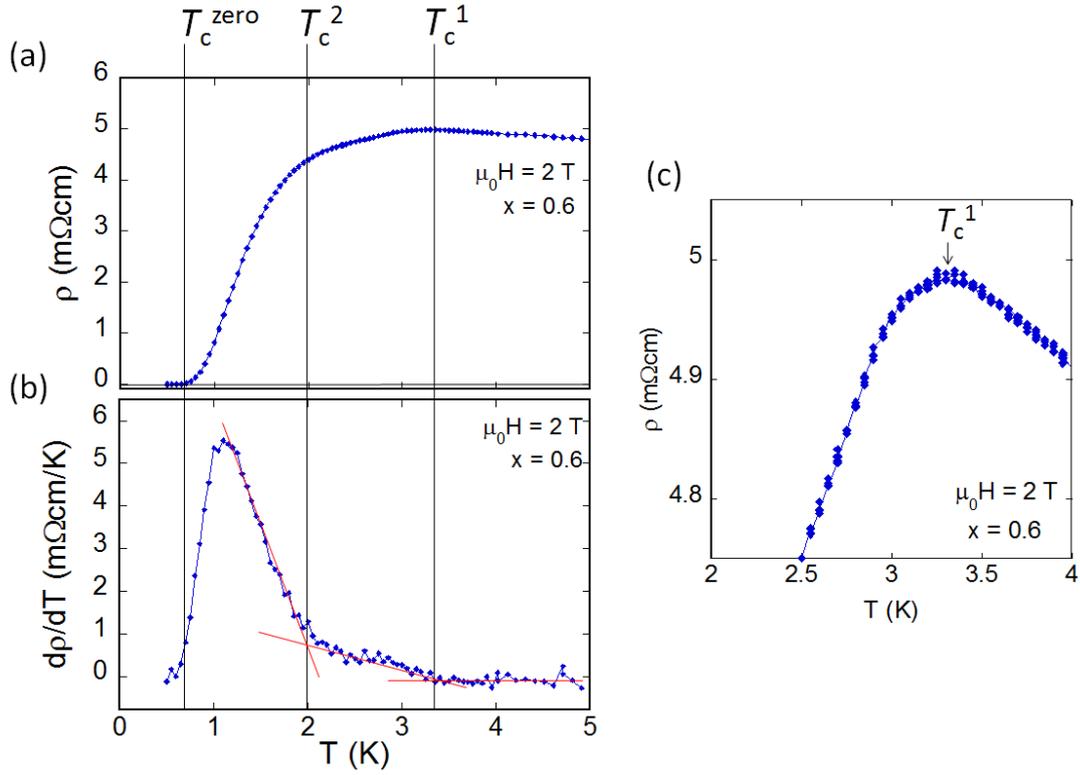

Fig. S10. (a) $T$ dependence of $\rho$ for $x = 0.6$ at 2 T. (b) $T$ dependence of d$\rho$/d$T$ for $x = 0.6$ at 2 T and estimation of $T_c^{zero}$, $T_c^2$, and $T_c^1$. (c) Enlarged $\rho$-$T$ plot at around $T_c^1$.